# Solar PV Frequency Control in the U.S. EI and ERCOT Interconnections — Case Studies and Recommendations

Shutang You
University of Tennessee, Knoxville
Knoxville, TN, U.S.
syou3@utk.edu

*Abstract* — This paper studied the solar PV frequency control in the U.S. Eastern Interconnection (EI) and Texas Interconnection (ERCOT) systems. The studied frequency control approaches include droop frequency control, inertia control, and droop-inertia-combined frequency control. The control effects of different frequency controls of PV in the EI and ERCOT are studied using actual high PV penetration interconnection grid models to provide suggestions to the future revision of future PV frequency control standards.

## I. Introduction

Due to the decreasing price, wind and photovoltaic (PV) generation penetration is increasing many power systems. Similar to other renewable generation, PV usually runs at the maximum power point, providing no frequency response to the power grid. The displacement of synchronous generators with PV has direct impacts on the system inertia level and frequency regulation capability. Many power systems starts to notice the risks of insufficient system inertia and frequency regulation resources due to the increase of PV and other non-synchronous renewable generation, largely thanks to the deployment of wide-area synchrophasor measurement systems.

As more solar PV generation is integrated into the power grid, frequency control using PV plants may become essential to ensure system frequency stability. Table 1 shows a comparison between various standards applicable in North America. It can be seen that the low-frequency and high-frequency ride through and frequency-watt functions have been gradually incorporated into the latest standards. The current U.S. power grids have standards on PV frequency droop control (primarily for overfrequency regulation) but no requirement on synthetic inertia control. As PV penetration increases, more detailed specifications on inertia control are expected to be included to enhance the system frequency control capability of PV.

As the future requirements on PV-based frequency control are not clear, this paper combines the simulation results of PV frequency droop control and inertia control in the U.S. Eastern Interconnection (EI) and the Texas Interconnection (ERCOT) systems with the current PV frequency control standards in North America to provide recommendations to future frequency control standards. The rest of this paper is organized as follows: Section II discussed the PV frequency droop control; Section III studied the PV inertia control; Section IV gives the conclusions.

## II. PV Frequency Droop Control

### A. Current Standard Review

PV frequency droop control (primarily for overfrequency regulation) has been included in multiple standards in North America, as shown in Table 1. The NERC reliability guideline on Bulk Power System (BPS)-connected inverter-based resource performance [1] and the latest IEEE 1547 standard [2] published in April 2018 require that smart inverters provide frequency-watt function to decrease real power to stabilize over-frequency events. If active power is available, inverters should also increase real power to support low frequency.

The NERC reliability guideline [1] specifies that "the active power-frequency control system should have an adjustable proportional droop characteristic with a default value of 5%". In addition, frequency droop should be based on the difference between maximum nameplate active power output ($P_{max}$) and zero output ($P_{min}$) such that the droop line is always constant for a resource. The NERC guideline [1] also has requirements on the dynamic active power-frequency performance. It specifies values of the reaction time, rise time, settling time, overshoot, and settling band after a 0.002 p.u. (0.12Hz) step change in frequency from nominal 60Hz. The IEEE 1547 standard requires droop frequency control for overfrequency events but does not explicitly require a specific droop rate.

Table 1. Comparison between various standards applicable in North America

| Function Set | Advanced Functions | Interconnection Standard | | | | State/PUC Rules | |
|---|---|---|---|---|---|---|---|
| | | IEEE 1547 - 2003 | IEEE 1547-2014 | IEEE 1547.1-2018 | IEEE P2800 (ongoing) | CA Rule 21-2015 | NREC "Reliability Guideline" (Sept. 2018) |
| Frequency Support | L/H Frequency Ride-Through | N/A | N/A | ☑☑ | ☑☑ | ☑☑ | ☑☑ |
| | Frequency-watt (Droop) | ☒ | ☑ | ☑☑ | ☑☑ | N/A | ☑☑ |
| | Synthetic Inertia | N/A | N/A | N/A | | N/A | N/A |
| ☒-Prohibited; ☑-Allowed by mutual agreement; ☑☑-Capability Required; | | | | | | | |

This work is funded in whole by the U.S. Department of Energy Solar Energy Technologies Office, under Award Number 30844. This work also made use of Engineering Research Center Shared Facilities supported by the Engineering Research Center Program of the National Science Foundation and the U.S. Department of Energy under NSF Award Number EEC-1041877 and the CURENT Industry Partnership Program.

## B. Simulation Results in the U.S. EI and ERCOT Systems

Fig. 1 shows the control diagram of frequency droop control using PV power plants. PV droop frequency control consists of three blocks: a frequency deadband, a low-pass filter, and a control gain. The functions of the low-pass filter and the control gain are to remove fast fluctuations and adjust the response magnitude, respectively. The control effects of droop frequency control in the 80% renewable penetration scenarios of the U.S. Eastern Interconnection (EI) and Texas Interconnection (ERCOT) are shown in Fig. 2 to Fig. 5. The blue curves show the cases without PV droop frequency control, while the yellow curves represent the cases with PV droop frequency control. It can be seen that droop frequency control can improve the setting frequency and frequency nadir for both grids.

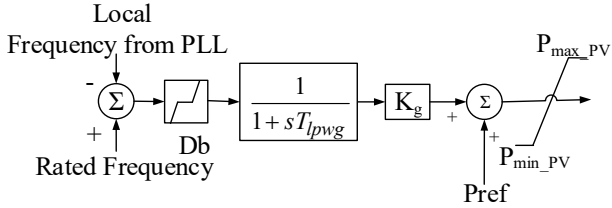

Fig. 1. Frequency droop control using PV

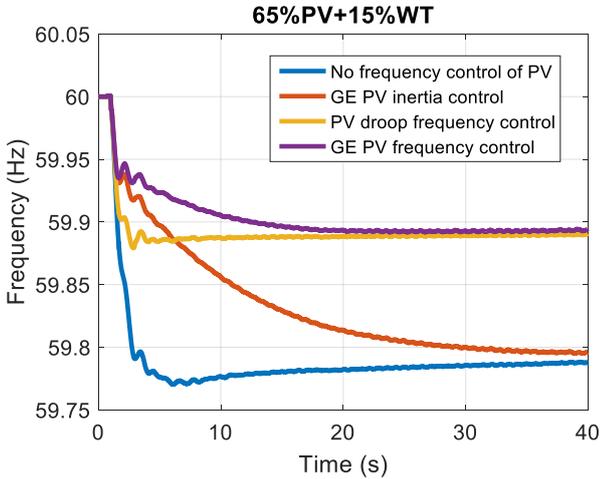

Fig. 2. System frequency response using different PV control strategies in the EI (80% renewable)

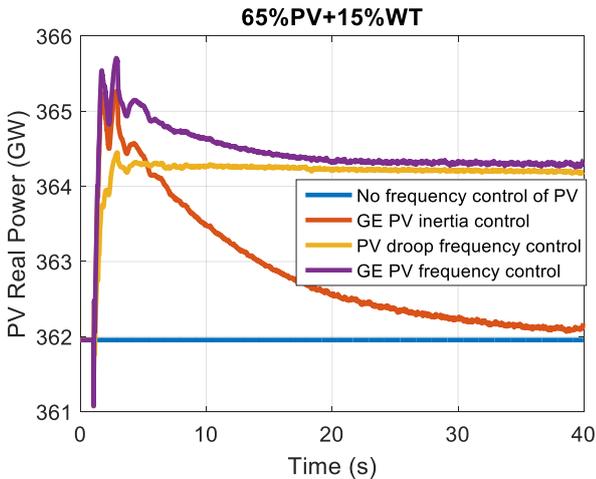

Fig. 3. PV output using different frequency control strategies in the EI (80% renewable)

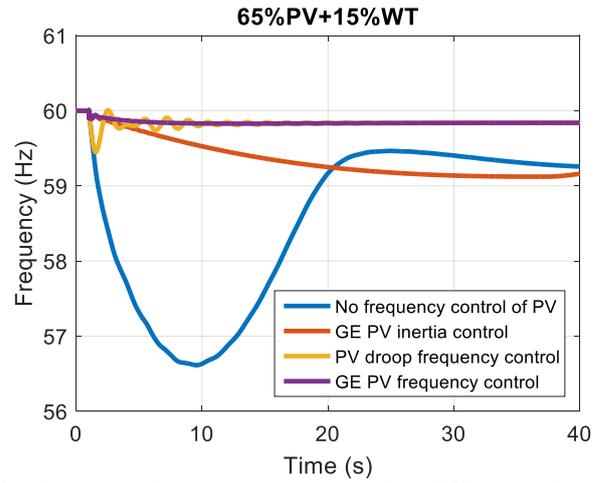

Fig. 4. System frequency response using different PV control strategies in the ERCOT (80% renewable)

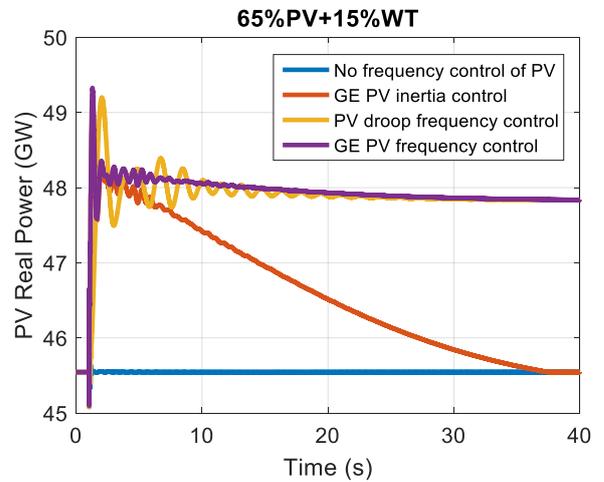

Fig. 5. PV output using different frequency control strategies in the ERCOT (80% renewable)

## C. Study Observations

Both the NERC guideline [1] and IEEE 1547 standard [2] recognize that renewable generators are typically operating at the maximum power points. Underfrequency response is not mandatory if there is no real power headroom available for frequency response. However, the simulation results of the U.S. EI and ERCOT interconnection grids under high PV penetration indicate that PV headroom reserve may be necessary for underfrequency response when the PV penetration is high and other frequency regulation resources, including fast load response and energy storage, are insufficient. Underfrequency events, which indicate insufficient generation, are also critical to system security in high renewable penetration scenarios.

The NERC guideline [1] requires an adjustable proportional droop characteristic with a default value of 5% and using the maximum nameplate active power output to determine a constant droop line. This specification makes it easier to predict the active power- frequency response. However, in high inverter penetration scenarios, this requirement is equivalent to having all inverter-based resources to have governor response characteristics for overfrequency events. This may be unnecessary. For example, the current EI system only has around 30% governor response generation and even less generation capacity that actually responds to frequency change. It may also be unnecessary for all inverter-based resources to reduce output during overfrequency periods, which may reduce their efficiency.

In addition, the NERC guideline specifications on the dynamic active power-frequency performance is assessed based on a specific step change of frequency [1]. Simulation results in the EI and ERCOT systems show that in some grid conditions, especially in high renewable penetration, inverters that meet these specifications may not necessarily satisfy the system frequency support requirement well. For example, the reaction time and the rise time may need to be significantly smaller to arrest the rapid frequency decline in a high renewable low inertia ERCOT system.

### D. Recommendations

The IEEE 1547 standard and the NERC guideline did not specify PV headroom requirements for underfrequency events. Headroom reserve for underfrequency frequency response may be necessary when inverter-based renewable penetration is high, especially for the ERCOT system in high-renewable low-inertia scenarios. A tradeoff between economics and reliability may be necessary when setting headroom reserve requirement for underfrequency event response. Since 5% as the default droop characteristic for all inverter-based resources in high inverter penetration scenarios may result in excessive frequency response and cost increase, it may be necessary to procure only some of inverter-based resources to provide this frequency-watt grid service. In some other scenarios, inverters may need to apply a smaller droop rate to provide stronger support to system frequency stability. System operators may need to determine the droop characteristic, the responsive resource portfolio, and the headroom reserve based on a detailed assessment on the real-time conditions of system and inverter-based resources.

As the system renewable penetration increases, the dynamic change of the system frequency response will become faster due to the inertia decrease. Therefore, the dynamic active power-frequency performance of PV frequency control needs to consider the grid dynamics. More detailed requirements should be determined based on detailed analysis of the grid event frequency measurements and extensive simulations.

## III. PV SYNTHETIC INERTIA CONTROL

### A. Literature Review

Synthetic inertia has three desired functions: keeping the initial ROCOF value within ride-through capabilities of generators and loads; delaying nadir time to earn time for primary frequency response, and limiting the frequency nadir to avoid load/generator disconnection (in some systems with very low inertia). While few studies have studied PV synthetic inertia control, most inverter inertia control studies focus on wind generation, electric vehicles or inverter-based energy storage, HVDC, and virtual synchronous generators (VSG). As of 2018, U.S. power grids have no standard or requirement on synthetic inertia control for inverter-based resources, as shown in Table 1.

Studies on inertia control usually use the inverter real power control loop to mimic the inertia response of conventional synchronous generators. High-pass filtered frequency signal and df/dt are the two common inputs for inertia control. For example, Ref. [3] studied virtual inertia emulation in wind turbine generator inverters and used the high-pass filtered frequency signal to control electrical output. Ref. [4] applied df/dt in control to provide inertia response from HVDC-interfaced wind power. In summary, existing studies on synthetic inertia have some findings as follows.

*1) The need for synthetic inertia for frequency regulation*

The need for synthetic inertia applies for small synchronous grids with high penetration of renewables and large synchronous grids to prevent total system collapse in case of a system split and subsequent islanding operation [5]. In some small synchronous grids, synthetic inertia from renewable generation is becoming a requirement. For example, in Hydro Quebec, wind generation resources (>10 MW) during underfrequency conditions are required to provide momentary overproduction that limits the frequency drop after a major loss of generation.

*2) Inertia response speed after the start of a contingency*

Ref. [6] found that the "synthetic inertia" fast frequency response (FFR) type devices have the potential to prevent high ROCOF events (ROCOF is high after the contingency), but the time period required to reliably detect and measure ROCOF events to ensure the appropriate response to mitigate the events poses some challenges. Ref. [7] found that the synthetic inertia needs to have a short reaction time and power ramp at a sufficient rate to satisfy several criteria to maintain ROCOF within a specific range.

*3) Inertia response characteristics during frequency recovery*

Studies found that a suitable form of synthetic inertia control is needed to prevent unintended adverse system issues during the frequency recovery [8]. Recognizing synthetic inertia may be unfavorable to frequency recovery, Ref. [9] found that compared with fixed inertia of conventional generators, emulating varying inertia may help improve the system frequency stability. A study in Ref. [10] showed that some virtual inertia controls that need to be paid back (for example, wind generation) may not always be beneficial to system frequency response, especially when the system governor response is fast.

*4) Inertia response power magnitude*

Ref. [7] found that a minimum response power is needed to constrain ROCOF. This study also found that synthetic inertia response is required for both overfrequency and underfrequency events.

*5) Stability impacts*

Rotational inertia is important for power grids to maintain stability after disturbances. However, studies show that synthetic inertia can improve frequency response but it may or may not improve small signal stability. Sometimes, synthetic inertia can amplify the instability. Ref. [11] found that inertia controllers need to be designed individually for each power grid to avoid instability caused by the limited bandwidth of df/dt.

### B. Simulation Results in the U.S. EI and ERCOT Systems

Fig. 6 shows the diagram of inertia control using PV power plants. The PV inertia controller includes a frequency deadband, a low-pass filter, a control gain, and a high pass filter. The frequency deadband and the low pass filter eliminate the high frequency noises. The high-pass filter calculated the derivative of the frequency to emulate inertia response. The inertia control can be combined with droop frequency control, as shown in Fig. 7. The simulation results of the U.S. EI and ERCOT systems using inertia control and inertia-droop-combined control are also shown in Fig. 2 to Fig. 5.

It can be seen that at the initial stage of the disturbance, the inertia power is dominant because of the large value of the rate of change of frequency (ROCOF). As the frequency decreases further, the virtual inertia power decreases due to the decrease of ROCOF, while PV droop frequency control response increases with the system frequency deviation. Therefore, the primary contribution of PV inertia control is to constrain the ROCOF, and delay the nadir occurrence time. This allows more time for some slow-reacting primary frequency response resources (such as turbine governors with relatively large time constants) to take effects to prevent under-frequency load shedding.

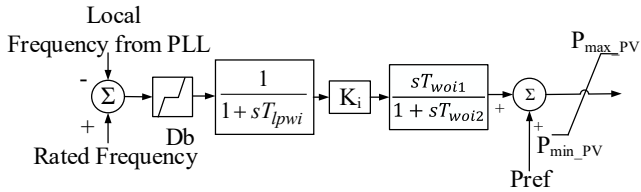

Fig. 6. GE synthetic inertia control using PV

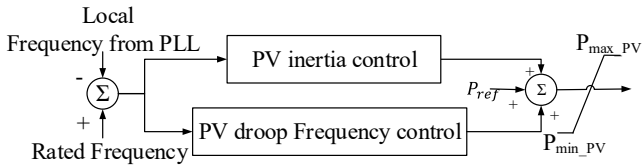

Fig. 7. Primary frequency control using PV

### C. Study Observations

*1) Synthetic inertia benefits frequency stability and the need for synthetic inertia depends on the original inertia level of the interconnection*

The study results show that for large synchronized grids such as the EI, droop frequency response from PV may be sufficient for frequency regulation due to the fast response of inverters and remaining inertia support in the system. If the system scale is large, adding synthetic inertia control to the droop frequency control can delay the frequency nadir time but not significantly change the frequency nadir. For the smaller ERCOT system, the benefit of inertia is more obvious in terms of improving frequency nadir compared with that of the EI.

*2) Synthetic inertia characteristics have direct impacts on frequency response*

The response characteristics of inertia control is found to have direct and significant impact on frequency response performance. Reaction time, the ramp rate, and the response magnitude are critical features for effective synthetic inertia control. In addition, the synthetic inertia response is more critical for underfrequency events, which require PV headroom reserve.

*3) Other resources that provide fast frequency response can provide alternatives of inertia response and alleviate the reliance on inertia*

Both EI and ERCOT can leverage other options to provide inertia response, including increasing inertia from synchronous generators by committing more units, deploying synchronous condensers, modifying some governor response characteristics of synchronous units to allow fast response, fast load response, energy storage, and many other inverter-based DERs.

### D. Recommendations

*1) Quantifying synthetic inertia requirement needs constant monitoring of system inertia, fast frequency response resources, and the largest possible contingency*

As renewable penetration increases, grid operators should monitor the grid inertia level and frequency response to assess the necessity of synthetic inertia from PV and avoid large-scale system collapse due to low inertia and resource contingencies. Because of its direct impact on fast frequency response, real-time information of on-line fast frequency response resources should also be considered in quantifying the demand on system inertia. Knowing the system inertia and frequency response conditions can also help to determine the optimal headroom reserve for frequency control. With this information, more efficient frequency control can be realized. Fig. 8 shows a schematic diagram of frequency control in future high renewable grids.

*2) Control solution to provide synthetic inertia needs detailed analysis considering the dynamics of specific grids*

The synthetic inertia control solution for a specific grid will require detailed analysis considering system dynamic characteristics. The analysis should consider changes in the inertia and governor response resources, measurement delay/errors and ambient noises, possible contingencies with different magnitude, and fault ride through, etc.

*3) Headroom reserve is required to provide effective synthetic inertia response from PV in underfrequency events*

As it is easier for inverter-based renewable generation to reduce power output than to increase output, frequency regulation for underfrequency events is more challenging. Therefore, synthetic inertia response during underfrequency events is more important for grid security. If other resources, for example, energy storage, are unavailable, inertia response from PV in underfrequency events requires headroom reserve and PV power curtailment. The amount of PV headroom to maintain the frequency nadir above a specific value is determined by the needed synthetic inertia quantity and ROCOF, which further depends on the system inertia level, the largest possible contingency magnitude, and the system primary frequency response resources.

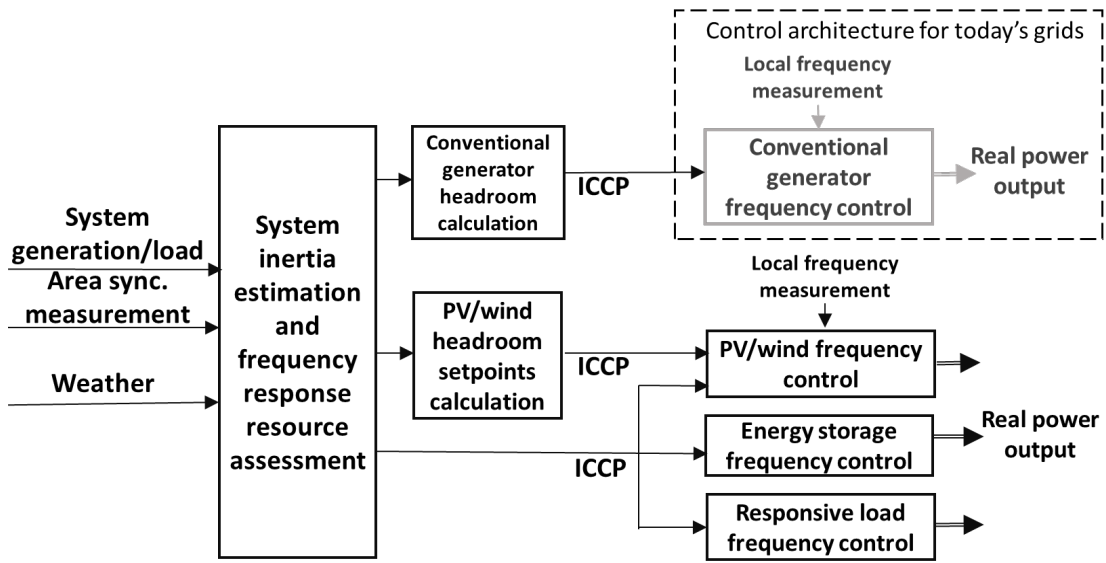

Fig. 8. The control architecture for future high renewable grids

*4) Response power from synthetic inertia is needed during frequency recovery*

In the frequency recovery stage, synthetic inertia response is preferred to be kept above zero (for underfrequency events) or below zero (for over-frequency events) to prevent adverse effects on the system frequency response. In contrast, synchronous inertia from conventional generators with governors does not have this issue because the power increased by governor response through steam valves and water flow is sufficient to cover both the rotor speed (kinetic energy) recovery and the generator real power output for frequency response, ensuring net increase in real power output (for underfrequency events) and net decrease (for over-frequency events).

## IV. Conclusions

This paper studied the effects of droop frequency control and inertia control in the U.S. EI and ERCOT power grids. The study results were analyzed to provide recommendations to future PV-based frequency control standards in the U.S. These study results and control suggestions can also be references for other power grids operating with high penetration of renewables.

## References

1. NERC, *BPS-Connected Inverter-Based Resource Performance.* 2018.
2. IEEE PES Industry Technical Support Task Force, *Impact of IEEE 1547 Standard on Smart Inverters.* 2018.
3. Arani, M.F.M. and E.F. El-Saadany, *Implementing virtual inertia in DFIG-based wind power generation.* IEEE Transactions on Power Systems, 2013. **28**(2): p. 1373-1384.
4. Miao, Z., et al., *Wind farms with HVdc delivery in inertial response and primary frequency control.* IEEE Transactions on Energy Conversion, 2010. **25**(4): p. 1171-1178.
5. ENTSO-E, *Need for synthetic inertia (SI) for frequency regulation.* 2018.
6. EirGrid and SONI, *RoCoF Alternative Solutions Technology Assessment - Project Phase I Study Report.* 2015.
7. EirGrid and SONI, *RoCoF Alternative & Complementary Solutions - Project Phase 2 Study Report.* 2016.
8. Morren, J., et al., *Wind turbines emulating inertia and supporting primary frequency control.* IEEE Transactions on power systems, 2006. **21**(1): p. 433-434.
9. Alipoor, J., Y. Miura, and T. Ise, *Power system stabilization using virtual synchronous generator with alternating moment of inertia.* IEEE Journal of Emerging and Selected Topics in Power Electronics, 2015. **3**(2): p. 451-458.
10. Van de Vyver, J., et al., *Droop control as an alternative inertial response strategy for the synthetic inertia on wind turbines.* IEEE Trans. Power Syst, 2016. **31**(2): p. 1129-1138.
11. Duckwitz, D. and B. Fischer, *Modeling and Design of df/dt-based Inertia Control for Power Converters.* IEEE Journal of Emerging and Selected Topics in Power Electronics, vol. preprint, 2017.